\documentclass[prl,aps,twocolumn,showpacs]{revtex4}
\usepackage{graphicx}
\usepackage{dcolumn}

\begin{document}
\def\be{\begin{equation}}
\def\ee{\end{equation}}
\def\bearr{\begin{eqnarray}}
\def\eearr{\end{eqnarray}}
\def\la{\langle}
\def\ra{\rangle}
\def\l{\left}
\def\r{\right}

\title{Bragg spectroscopy of discrete axial quasiparticle modes in a 
cigar-shaped degenerate Bose gas}
\author{Tarun Kanti Ghosh}
\affiliation
{ The Abdus Salam International Center for Theoretical Physics, Strada Costiera 11,
34014 Trieste, Italy.}

\date{\today}

\begin{abstract}
We propose an experiment in which  long wavelength discrete axial quasiparticle modes 
can be imprinted in a 3D cigar-shaped Bose-Einstein condensate by using 
two-photon Bragg scattering experiments,
similar to the experiment at 
the Weizmann Institute [J. Steinhauer {\em et al.}, Phys. Rev. Lett. {\bf 90}, 060404 (2003)] 
where short wavelength axial
phonons with different number of  radial modes have been observed. 
We provide values of the momentum, energy and time duration of the two-photon 
Bragg pulse and also the two-body interaction strength which are needed in the Bragg 
scattering experiments in order to observe 
the long wavelength discrete axial modes.
These discrete axial modes can be observed when the system is dilute and the time
duration of the Bragg pulse is long enough. 
\end{abstract}


\maketitle

Bose-Einstein condensates (BEC) \cite{rmp} of alkali-atoms represent a wonderful testing ground
of theories of weakly interacting bosons. The atoms are spatially confined by a harmonic
trap and inhomogeneity makes these systems even more interesting. By changing the trapping
frequency one can produce quasi-1D, quasi-2D as well as cigar-shaped but 3D Bose systems. 
Due to the reduction in the dimension and different shape, the effect of density and phase 
fluctuations become prominent which produces many interesting features.
Bragg spectroscopy \cite{brag,stamp} of a trapped BEC has become an important tool to reveal many 
bulk properties such as dynamic structure factor, verification of the Bogoliubov excitation spectrum
\cite{ste,davidson}, observation of the Bogoliubov quasiparticle amplitudes \cite{bru},
momentum distribution and correlation functions of a phase fluctuating quasi-1D Bose systems
\cite{ric,gerbier}.
Bragg spectroscopy has also been used for tuning and measuring the scattering length of atoms
by using the optical Feshbach resonance \cite{theis}.

In the Bragg scattering experiments, 
the condensate is excited by using two Bragg pulses with approximately parallel 
polarization, separated by an angle $\theta$. The pulses have a frequency difference $\omega $ 
determined by two acousto-optic modulators. If a photon is absorbed from the higher-frequency
($\omega_h $) beam and
emitted into other ($\omega_l $), an excitation is produced with energy 
$\hbar \omega = \hbar (\omega_h - \omega_l)$ and momentum
$ \hbar \bf k $, where $k = |{\bf k} | = 2k_p sin(\theta/2)$, and $k_p$ is the photon wave 
number.
The wave-vector $ \bf k$ is adjusted to be along the $z$-axis. 
Both the values of $ k$ and $ \omega $ can be tuned by changing the angle between two beams and
varying their frequency difference. For small values of $k$, the system is excited in the
phonon regime and the response is detected by measuring the net momentum imparted
to the system. 
Note that, all the momentum response experiments \cite{stamp,ste,davidson} are limited to at
most a quarter of radial trapping period $T_0 $ ($T_0 =2 \pi /\omega_0 $) and the energy 
transfer $\omega$ is chosen to be the order of kHz so that $ \omega >\omega_0 $
(where $ \omega_0 $ is the radial trapping frequency) which excites the different 
 radial modes. In those experiments \cite{stamp,ste}, 
the local density approximation has been used to the actual inhomogeneous condensates, 
by using the Bogoliubov theory of uniform gases.  

The axial excitations of a cigar shaped BEC can be divided into 
two regimes: short wavelength excitations whose wavelength is 
much smaller than the axial size and long wavelength excitations whose 
wavelength is equal or larger than the axial size of the system. 
The short wavelength axial phonons with different number of radial 
modes of a cigar-shaped condensates which give rise 
to the multibranch spectrum \cite{zaremba} has been resolved in a 
Bragg spectroscopy with a long duration ($t_B > T_0 $) of the Bragg pulses 
\cite{davidson,tozo}. 
In this experiment the condensate is excited by the Bragg 
pulses and allow it for free expansion. During the expansion 
of the condensate, these short wavelength
excitations decouple from the condensate itself, forming a separate cloud of atoms. The
total momentum is measured by counting the atoms which is decoupled from the condensate.

We consider a 3D Bose gas confined in a cigar-shaped harmonic 
trap. As described in Ref. \cite{davidson}, the condensate consists 
of $ N=10^5$ atoms of $^{87}$Rb. The radial and axial trapping frequencies are 
$\omega_0 =2 \pi \times 220$ Hz and $\omega_z =2 \pi \times 25$ Hz 
($a_z = \sqrt{\hbar/m \omega_z} = 2.155 \mu m$), respectively.
Recently, the low-energy axial modes of a 3D BEC has been studied 
theoretically \cite{tozzo1} and experimentally  \cite{katz} by using
a matter-wave interference technique.
In those works \cite{tozzo1,katz}, they
have shown that the long wavelength phonon do not form a separate 
cloud during the expansion. At long expansion times and for a given $ k $, the excitations
can separate out from the condensate only if
$ k a_0 > k_c a_0 = \pi (\omega_z/\omega_0) (\eta/2)^{1/2} $, while for $ k < k_c $, they
remain within the condensate at all times \cite{tozzo1}. Here, $ \eta = \mu/\hbar \omega_0 $ and 
$ \mu $ is the chemical potential.
Therefore, in order to observe the long wavelength discrete axial modes by using
the Bragg spectroscopy as in Ref. \cite{davidson}, we have to choose the parameters such that 
the low energy excitations can separate out from the condensate for low value
of $ k $. This is an alternative way to realize the low-energy discrete axial modes in contrast 
to the matter-wave interference technique \cite{katz}. 

In order to get the separate cloud of low-energy axial modes, the
chemical potential $ \eta $ must be small.
One can reduce the two-body scattering length $a$ by using the Feshbach resonance \cite{theis} so
that the axial size decreases and the long wavelength 
excitations can separate out easily  from the condensate. 
If we take $ a = 13.63 \times 10^{-11} m$, instead of the bare scattering length for Rubidium and 
other parameters are remain fixed as it is in Ref. \cite{davidson},
then $ \mu = 2 \hbar \omega_0 $. 
Of course, one can also get the chemical potential $ \mu = 2 \hbar \omega_0 $
by reducing the number of atoms in the condensate while keeping the bare value
of the two-body scattering length.
For this choice of $ a$, $ k_c a_0 = 0.35 $ and 
we will show that the long wavelength excitations can be separated out of the condensate. 
Note that for this choice of scattering length, the system is still being 3D.
The radial and axial sizes of the condensate are
$ R_0 \sim 1.46 \mu m $ and $Z_0 \sim 12.78 \mu m $, respectively,
so that it becomes a cigar-shaped condensate with the deformation parameter 
$ \lambda = \omega_0/\omega_z \sim 9$. 

Note that in the experiment \cite{davidson}, the energy transfer 
$\omega$ is chosen to be the order of kHz ($\omega > \omega_0 $) 
which excites the short wavelength axial phonons with different radial modes. Since the 
wave-lengths of these
excitations are very small compared to the axial size and large compared to 
the radial size, then the axial quasiparticle modes can be described by the plane-wave states
and the radial excitations can be described by the discrete modes which are the 
manifestation of the inhomogeneity along the radial direction. Therefore, the short wavelength phonons 
propagate 
along the $z$-axis with different number of 
radial modes which have been observed in Ref. \cite{davidson}. 
If $\omega$ is the order of $ \omega_z $ but less than the radial trapping frequency $\omega_0$
i.e. $\omega_z \leq \omega << \omega_0$, and the wave-vector of the Bragg pulse is comparable to 
the inverse of the axial size
then it would excite only the different  number of long wavelength axial modes, instead of 
short wavelength axial phonons with different numbers of radial modes. Therefore, the wave-length of 
these excitations are comparable to the
axial size and the effect of inhomogeneity along the axial direction has to be included in the 
studies. The inhomogeneity along the axial direction would be manifested by the presence of the discrete 
axial modes. 
The discrete long wavelength axial modes due to the finite size of 
the axial direction can be observed by measuring the dynamic structure factor, $S(k,\omega)$, 
which is related to the momentum transferred $P_z(t)$ due to the low-energy two-photon 
Bragg scattering.

Within the Thomas-Fermi approximation ( i. e. neglecting the
quantum pressure term), the hydrodynamic description  
for the axial density fluctuations $\delta n(z)$
of a 3D cigar shaped Bose system has been studied in \cite{stringari}. 
In the dimensionless form, the equation for the $\delta n(z)$ reads as
\be \label{main}
[(1- \tilde z^2)\nabla_{\tilde z}^2 -4 \tilde z \nabla_{\tilde z} + 
4 \tilde \omega^2] \delta n(\tilde z) = 0,
\ee 
where $ \tilde z = z/Z_0 $ and $ \tilde \omega = \omega/\omega_z $.
The eigenvalues of the above equation are $(\omega_j/\omega_z)^2 = j(j+3)/4$ and the corresponding
normalized eigen functions are
\be
\delta n(\tilde z) \sim \psi_j(\tilde z) = \sqrt{\frac{(j+2)(2j+3)}{8(j+1) \pi R_{0}^2 Z_0}}
P_{j}^{(1,1)} \l (\tilde z \r ),
\ee
where $ P_{j}^{(1,1)}(\tilde z)$ is the Jacobi polynomial. 
Eq. (\ref{main}), including the above eigenvalues and the eigen vectors
are valid only when the system parameter satisfies the condition:
$ \mu >> \hbar \omega_0 >>  \hbar \omega_z $.
In our method, the chemical potential is not too large compared to
the first radial excitation, but it is quite large compared to the
first axial excitation. Therefore, 
the Thomas-Fermi approximation is a reasonable approximation, since we are
dealing with the axial modes only and there is no coupling with the
radial modes.
The quantum numbers $j=1 $ and $j=2$ correspond to the axial center-of-mass and the
breathing modes with the frequencies $ \omega_1 = \omega_z $ and $ \omega_2 = \sqrt{5/2} \omega_z $,
respectively. Only these two frequencies have been measured with high accuracy by using the time-dependent
modulation of the trapping potential \cite{breathing}. It is very difficult to measure 
the frequencies of the other axial modes of large
quantum numbers by using the time-dependent modulation of the trapping potential and
therefore there are no measurements of the frequencies of these higher modes. 
It is useful to verify the frequencies of the other axial modes ($j>2$) to make
sure that the energy eigen values derived from the hydrodynamic approximation
are correct.  These modes can be easily 
verified by 
using the Bragg spectroscopy which is also our main concern of this work. 

The dynamic structure factor is obtained from the Fourier transform of the time-dependent
density-density correlation functions,
\be
S(k,\omega)  = \int dt \int dz e^{i(\omega t - kz)} < \delta \hat n(z,t) \delta \hat n(0,0)>,
\ee
where the density fluctuation operator is
\be
\delta \hat n(\tilde z,t) = \sum_j i\ \sqrt{\frac{\hbar \omega_j}{2 g}} 
\psi_j (\tilde z ) e^{-i \omega_j t} \hat \alpha_j + H.c.
\ee
Here, $ g = 4\pi a \hbar^2/m $ is the interaction strength determined by the $s$-wave 
scattering length $a$ and $ \hat \alpha_j $ is the 
quasi-particle operator of the $j$-th mode.
It is the density fluctuation spectrum that can be measured in the two-photon Bragg
spectroscopy.
At $T=0$, the dynamic structure factor can be written as
\be \label{dynamic}
S(k,\omega) = 
\sum_j A_j |\psi_j(k)|^2 \delta (\omega -\omega_j),
\ee
where $ A_j = (\frac{R_0^2 Z_0}{128 \pi^2 a a_z^2}) \frac{\sqrt{j(j+3)} (j+2)(2j+3)}{(j+1)} $ and 
$ \psi_j(\tilde k) = \int_{-1}^{1} d \tilde z e^{-i\tilde k \tilde z} P_{j}^{(1,1)}(\tilde z)$ 
is the spatial Fourier
transform of $ P_{j}^{(1,1)}(\tilde z) $. Here, $\tilde k = k Z_0 $ is the dimensionless wave vector.
We rewrite the dynamic structure factor as
$S( \tilde k,\omega) =  \sum_{j} S_j(\tilde k) \delta(\omega - \omega_j) $,
where $ S_j(\tilde k) = A_j |\psi_j(k)|^2 $ is the weight factor
which determines the weight of the Bragg-scattering 
cross-section in $S(k,\omega)$ of the corresponding axial modes of energy $ \hbar \omega_j$.
In Fig.1 we show $ S_j(\tilde k) $ as a function
of the dimensionless wave vector $ \tilde k = kZ_0 $ for the excitations $ j = 1,2,$ and $3$.
Fig.1 shows how many axial modes significantly contribute to $S(k,\omega)$ for a given momentum 
$\hbar k$. It is clear from Fig.1 that the strongest weights for these collective modes appear 
for $ kZ_0 \geq 2 $.
\begin{figure}[ht]
\includegraphics[width=8.5cm]{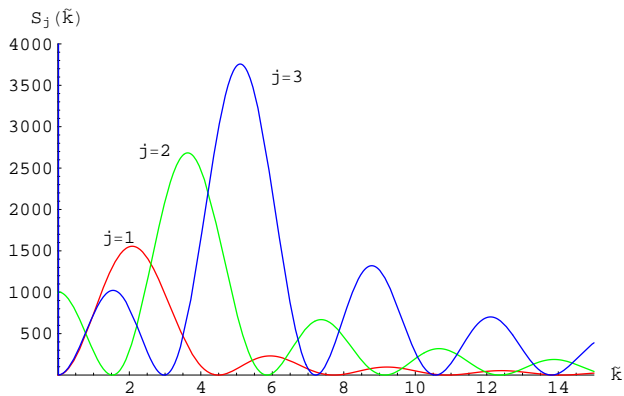}
\caption{(Color online) Plots of the weight factor $ S_j(\tilde k) $ vs the 
dimensionless wave-vector $\tilde k $ for $j=1,2,$ and 3.}
\end{figure}
As an example, for our choice of parameters
which gives $ Z_0 \sim 5.932 a_z $, this means that the momentum transfer $\hbar k$ in a
Bragg scattering experiments should be $ k \geq 0.337 a_z^{-1} $ for $ N = 10^5 $
in order to pick up the strong spectral weight from the low-energy collective modes.
However, in order to get the separate cloud of the low-energy excited atoms from the condensate, 
the wave-vector of the Bragg pulse must be $ k \geq 0.35 a_0^{-1} $ which implies that
the wave vector of the Bragg pulse must satisfy the condition: $ k Z_0 \geq 6 $.
 
We compare the dynamic structure factor given in Eq. (\ref{dynamic}) with the 
dynamic structure factor calculated within the local-density approximation (LDA).  
In the LDA, the dynamic structure factor is given by the analytic
expression \cite{stamp,zambeli}
\be
S_{LDA}( {\bf k}, \omega) = \frac{15 \hbar^2}{8} \frac{(\hbar^2 \omega^2 - E_{re}^2)}{E_{re} \mu^2} 
\l [ 1 - \frac{(\hbar^2 \omega^2 - E_{re}^2)}{2E_{re} \mu } \r ]^{\frac{1}{2}}.
\ee
Here, $E_{re} = \frac{\hbar^2 k^2}{2m}$ is the recoil energy and 
$ \mu = 0.5 \hbar \omega_z (15  \lambda^2 Na / a_z)^{2/5} $ is the chemical
potential. 
This analytic expression can also be used for 3D cigar-shaped trapped Bose systems.
The above expression can be recast for this system as,
\be \label{lda}
S_{LDA}( {\bf k}, \omega) = \frac{15}{2 \omega_z C^2} [ 2(\frac{\tilde \omega}{\tilde k})^2- 
\frac{\tilde k^2}{2C^4}] \l [1-2(\frac{\tilde \omega}{\tilde k})^2 + \frac{\tilde k^2}{2C^4} \r 
]^{\frac{1}{2}},
\ee
where $C= (15 \lambda^2 Na /a_z)^{1/5} \sim 5.93 $ for the experiment 
\cite{davidson} and $ \tilde \omega = \omega /\omega_z $.
In Fig.2 we plot the dynamic structure factor $S( \tilde k,\omega)$ vs the 
frequency $\omega $ by using the LDA 
as well
as by using the Fourier transformation of the density-density correlation function given in
Eq. (\ref{dynamic}).
For finite-energy resolution we have replaced the delta function in Eq. (\ref{dynamic}) by the 
Lorentzian with a width of $ \Gamma = 0.1 \omega_z$.

\begin{figure}[ht]
\includegraphics[width=8.5cm]{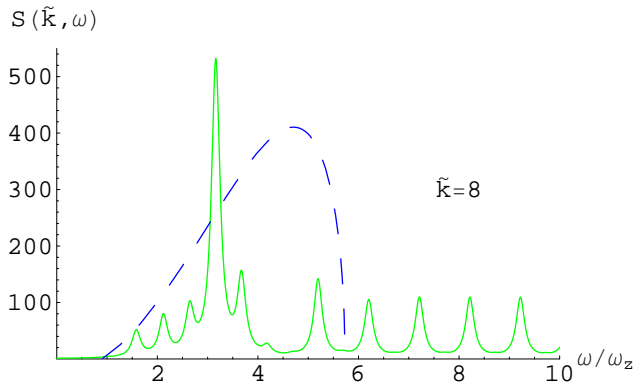}
\caption{(Color online) Plots of the dynamic structure factor $ S(\tilde k,\omega)$ vs normalized 
frequency $\omega/\omega_z$ at $T=0$.
The dashed lines shows the dynamic structure factor based on the LDA.}
\end{figure}

As one can see from Fig.2, the dynamic structure factor has multiple peaks. This phenomena is due to
the underlying discrete axial spectrum. Fig. 2 shows that for the given $ \tilde k = 8$, the most
prominent peaks at $ \omega = 3.162 \omega_z $ which corresponds to $ j=5 $ and other small peaks
corresponds to the other quantum numbers. In order to observe the $ j=5 $ quasiparticle modes, the wave
number of the Bragg pulse should be $ \tilde k = 8.0 $ which can be obtained by suitable choices of the
photon wave number ($k_p$) and the angle ($\theta $) between the Bragg beams. Note that, $ \tilde k =
8.0 $ implies $ k a_0 \sim 0.45 > k_c a_0 $. Therefore, the excited states will be separated out from
the condensate and the momentum measurements can be performed.  The simplified LDA picture fails to
produce the discrete structure of the axial modes, and it describes poorly the envelope of the spectrum
of the dynamical structure factor. However, the single peak in $ S_{LDA} $ occurs at $\omega_m \sim
4.75 $ for $ \tilde k = 8$, but the locations of the highest peak in $ S(k,\omega)$ is $\omega/\omega_z
= 3.162 $.  Therefore, the highest peak in $ S(k,\omega)$ do not match with the peak in the $ S_{LDA}
(k,\omega)$. The LDA fails to describe the discrete structure of the modes because it assumes the
system is locally uniform and it does consider partially the effect of the finite axial size.  We have
also carried out the same analysis for higher values of $k$. For $ \tilde k = 9 $, the condensate
responds resonantly at the frequency $ \omega = 3.674 \omega_z $ which corresponds to the axial quantum
number $j=6 $. In both the cases, the resonantly excited states are in the phonon regime since $ k \xi
<<1 $, where $ \xi = \sqrt{\hbar^2/2m \mu} \sim 0.168 a_z $ is the healing length.

The behavior of these multiple peaks in the dynamic structure factor can be resolved 
in two-photon Bragg spectroscopy, as
shown by Steinhauer {\em et al.} \cite{davidson}.
In the two-photon Bragg spectroscopy, the dynamic structure factor can not be measured 
directly. Actually, the observable in the Bragg scattering experiments is the momentum 
transferred to the condensate which is related to the dynamic structure factor and reflects 
the behavior of the quasiparticle energy spectrum.
The populations in the quasiparticle states can be controlled by using
the two-photon Bragg pulse. When the condensate is irradiated by an external moving optical
potential $ V_{op} = V_B(t) cos(kz-\omega t)$, the excited states are populated by the quasiparticle
with energy $ \hbar \omega $ and the momentum $\hbar k$, depending on the value of $k$ and 
$\omega$ of the optical potential $ V_{op} $. Here, $ V_B $ is the intensity of the
Bragg pulse. 
Suppose the system is subjected to a time-dependent Bragg pulse which is switched on at time 
$t>0$ and $k$ is also along the $z$-direction.
The momentum transfer to the Bose system from the optical potential can be calculated analytically either 
by using the Bogoliubov transformation \cite{blak} or by using the phase-density representations 
of the bosonic
order parameter \cite{tkg} and it is given by 
\bearr
P_z(t) & = & \sum_{j,k} \hbar k < \hat \alpha_j^{\dag} (t) \hat \alpha_j (t) >
 =  \l (\frac{V_B(t)}{2 \hbar} \r )^2 \sum_j \hbar k S_j (\tilde k) \nonumber \\
& \times & F_j [(\omega_j - \omega),t] - F_j [(\omega_j + \omega),t],
\eearr
where $ \hat \alpha_j (t) $ is the time-evolution of the quasiparticle operator of energy 
$\hbar \omega_j $ and  $ F_j [(\omega_j \pm \omega),t] = 
\l (\frac{sin[(\omega_j \pm \omega)t/2]}{(\omega_j \pm \omega)/2} \r )^2 $.
For positive $ \omega $ and a given $ \tilde k $ such that $ S_j(\tilde k) $ is maximum, 
the momentum transferred $ P_z(t)$ is resonant at the frequencies $ \omega = \omega_j $.
The width of the each peak goes like $ 2 \pi/t $. In order to resolve the different
peaks, the duration of the Bragg pulses should be at least of the order of the axial 
trapping period $ T_z = 2 \pi/\omega_z $. 
Moreover, the duration of the Bragg pulse, $t_B$, (atom-light interaction time) is a main 
factor in the Bragg spectroscopy and it
must be of order $ t_B > m/ \hbar k^2 $ in order to populate the Bragg reflected beam 
significantly \cite{keller}.
For large $t$ and $ \omega_z << \omega_0 $, $  P_z(t) \sim  S(k,\omega) $ \cite{tozo}. 
In Fig.3, we plot the net momentum transfer $P_z(t)$ vs the frequency 
$ \omega $ for three different choices of the time duration of the Bragg pulse. 
Fig. 3 shows that the shape of the $P_z(t)$ strongly depends on the time duration
of the Bragg pulse $ t_B$. 
When $ t_B = 1.1 T_z $, the $P_z(t)$ is a smooth curve with a single peak at 
$\omega/\omega_z = 1.8$, where $ T_z = 2\pi /\omega_z $ is the axial trapping period
which is 40 msec for the experiment \cite{davidson}.
When $t_B = 1.9 T_z $, there is a little evidence of few small peaks start developing in the
$P_z(t)$. 
When $ t_B = 3 T_z $, the multiple peaks in the $P_z(t)$ appears prominently.
The location of the peaks in Fig.3 for $ t_B = 3 T_z $ are exactly same as in Fig.2.
It implies that $P_z(t) \sim S((k,\omega)$ for a long duration of the Bragg pulse. 
Therefore, the multiple peaks in $S(k,\omega)$  are 
resolved in Fig.3 only when the duration of the Bragg pulse is $ t_B >> T_z $.
It should be noted that it is very difficult to use long duration of the Bragg pulse at the
present situation. It can induce a dipole oscillation of the whole condensate
in the trap  \cite{brun} and the reflection of phonons at the boundaries which might causes
a broadening of the response. 
We have checked that the $P_z(t)$ reflects the dynamic structure factor calculated from
the local density approximation if $ t_B < T_z $.
We also found that Fig.3 and the locations of the discrete peaks does not depend
on the intensity $V_B$ of the optical potential.
\begin{figure}[ht]
\includegraphics[width=8.5cm]{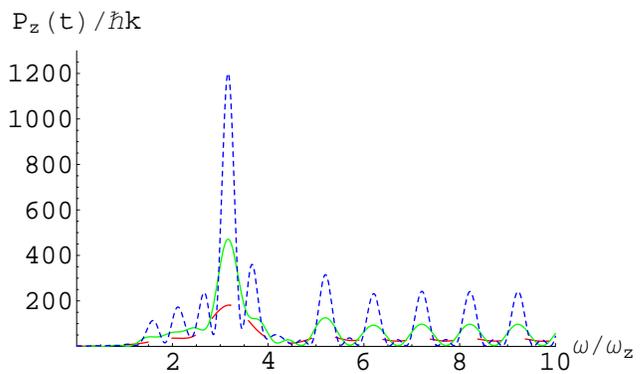}
\caption{(Color online) Plots of the net momentum transferred $ P_z(t)$ vs normalized frequency 
$ \omega /\omega_z $ at $T=0$ when
the dimensionless wave-vector is $ \tilde k = 8 $ for various time duration of the
Bragg pulse: $ t_B = 1.1 T_z $ (dashed), $ t_B = 1.9 T_z $ (solid)
and $ t_B = 3.0 T_z $ (dotted). Also, we have assumed $ V_B = 0.05 \hbar \omega_z$.}
\end{figure}
We have also studied $P_z(t) $ for $\tilde k = 9$ and find that 
there is a large peak at $\omega = 3.674 \omega_z $ corresponds to the quantum number $j=6 $
if the time duration of the Bragg pulse is longer than the axial trapping period.
This peak is also occurred in the dynamic structure factor as we have already discussed.
 
One can do the same analysis for other values of $k$ as long as only the low-energy
axial excited states contribute significantly in the dynamic structure factor and 
these low-energy axial states should be less than the first radial excitation.
It would be interesting to study intermediate regimes of the energy and wave-vector 
where both the radial and axial discrete states are excited simultaneously and it 
would provide a rich physics of the dynamic
structure factor as well as the momentum transfer to the Bose system, due to the non-trivial 
structure of spectrum of the density fluctuations and its corresponding eigenfunctions.

In conclusion, we have proposed an experiment in which the long wavelength discretized axial 
modes in a cigar-shaped (but 3D) condensate can be imprinted by using the Bragg scattering experiments, 
similar to the recent experiment \cite{davidson} where the short wavelength axial phonons with 
discrete radial modes have been observed. We have estimated the two-body scattering length
$a$, also the  
values of the momentum ($\hbar k$) and energy ($\hbar \omega $) of the two-photon Bragg pulses 
which are needed in the Bragg scattering experiments in order to observe the low energy discrete axial 
modes. 
These discrete axial modes can be observed when the system is dilute enough and 
the time duration of the Bragg pulse is long i.e. $ 
t_B >> T_z $.

Present Address: Department of Physics, Okayama University, Okayama 700-8530, Japan.
E-mail:tkghosh@mp.okayama-u.ac.jp

\end{document}